\documentclass[draft,a4paper]{aipproc}

\layoutstyle{6x9}

\begin{document}
\thispagestyle{empty}
\begin{flushright}
LU TP 04-30\\
hep-ph/0409068\\
September 2004
\end{flushright}
\vskip3cm
\begin{center}

\section{Chiral Meson Physics at Two Loops$^*$}
\vskip2cm
{\bf Johan Bijnens}\\[1cm]
{Department of Theoretical Physics 2, Lund University,\\
           S\"olvegatan 14A, SE 22362 Lund, Sweden}
\vfill
{\bf Abstract}
\end{center}
An overview of Chiral Perturbation Theory calculations
in the mesonic sector at the two Loop level is given.
Discussed in some detail are the partially quenched case relevant for
lattice QCD, the general fitting procedures and $\pi\pi$,$\pi K$ scattering
as well as the determination of $V_{us}$ and $K_{\ell3}$ decays. 
\vfill
\noindent\rule{0.9\textwidth}{0.5pt}\\
{\footnotesize $^*$ Invited plenary talk presented at the 19th European Few-Body conference,
Groningen, The Netherlands, August 23-27, 2004.}
\setcounter{page}{0}
\newpage

\title{Chiral Meson Physics at Two Loops}

\author{Johan Bijnens}{
  address={Department of Theoretical Physics 2, Lund University,\\
           S\"olvegatan 14A, SE 22362 Lund, Sweden}
}

\begin{abstract}
An overview of Chiral Perturbation Theory calculations
in the mesonic sector at the two Loop level is given.
Discussed in some detail are the partially quenched case relevant for
lattice QCD, the general fitting procedures and $\pi\pi$,$\pi K$ scattering
as well as the determination of $V_{us}$ and $K_{\ell3}$ decays. 
\end{abstract}

\maketitle


\section{Introduction}

In this meeting we have talked mainly about nuclei and baryons but I will
concentrate on mesons. The reason is that the lightest mesons
are the simplest bound states present in Quantum Chromo Dynamics
(QCD). They have the minimal number of constituents and are the lightest state
so should be spatially the simplest. Their properties are to a large
extent determined by Chiral Symmetry which enforces vanishing masses
in the limit of zero current quark masses, the chiral limit,
as well as vanishing interactions
in the zero momentum and chiral limit. 
It is the combination of these two properties
that allows us to produce a well defined low energy theory, Chiral Perturbation
Theory (ChPT), as a consistent approximation to QCD. In the remainder I review
ChPT at two loops.

\subsection{(Effective) Field Theory, ChPT and lattice QCD}

The underlying idea is really the essence of (most of) physics. {\em Use
the right degrees of freedom}. When there is a gap in the
spectrum with a consequent separation of scales, we can build the
theory containing only the lighter degrees of freedom and include the effects
of the neglected high mass/energy states perturbatively by building
the most general (local) Lagrangian with the low mass degrees of freedom.
This leads in general to an infinite number of parameters so no predictivity
is left. But when these terms can be ordered in importance
by some principle, usually referred to as power counting, we have a finite
number of parameters at any given order and thus an effective theory.

We need to use field theory since it is the only known way to locally
combine quantum mechanics and special relativity. A Taylor expansion
cannot be used because in the chiral limit the continuum of
states gives it zero convergence radius.
As important is the fact that off-shell effects are fully under control.
The freedom allowed by this is fully described by extra free parameters.
In addition it is systematic, {\em all} effects at a given order in the
expansion are included and errors can be estimated. Drawbacks are of
course the large number of free parameters, but do not forget the fact that
while a model can have few parameters there is a large freedom in the
space of possible models hidden behind it, and as always the expansion
itself might also not converge.

For ChPT the included degrees of freedom are the Goldstone Bosons
from spontaneous breakdown of chiral symmetry, identified with
the pseudoscalar octet, $\pi^{\pm,0}$, $K^\pm$,$K^0$, $\overline{K^0}$ 
and $\eta$. The powercounting principle is dimensional
counting, counting powers of generic momenta $p$,
 and the expected breakdown scale is the mass of the neglected
resonances, on the order of $M_\rho$. 

Chiral symmetry is the interchange of $u,d,s$ quarks
which in the limit $m_q\to 0$ can be done independently for the
left and right chirality since the QCD Lagrangian,
\begin{equation}
{\cal L}_{QCD} =  \sum_{q=u,d,s}
\left[i \bar q_L D\hskip-1.3ex/\, q_L +i \bar q_R D\hskip-1.3ex/\, q_R
- m_q\left(\bar q_R q_L + \bar q_L q_R \right)
\right]\,,
\end{equation}
does not couple left and right in the chiral limit.
The symmetry group is $SU(3)_L \times SU(3)_R$ and is spontaneously broken
to the diagonal subgroup by
$\langle \bar q q\rangle = \langle \bar q_L q_R+\bar q_R q_L\rangle
\ne 0$. The 8 broken generators correspond to the pseudoscalart octet.
ChPT in its present form was introduced by Weinberg, Gasser and Leutwyler
\cite{Weinberg,GL1,GL2} and introductory lectures can be found
in \cite{ChPTlectures}.

\begin{figure}
\includegraphics[width=0.8\textwidth]{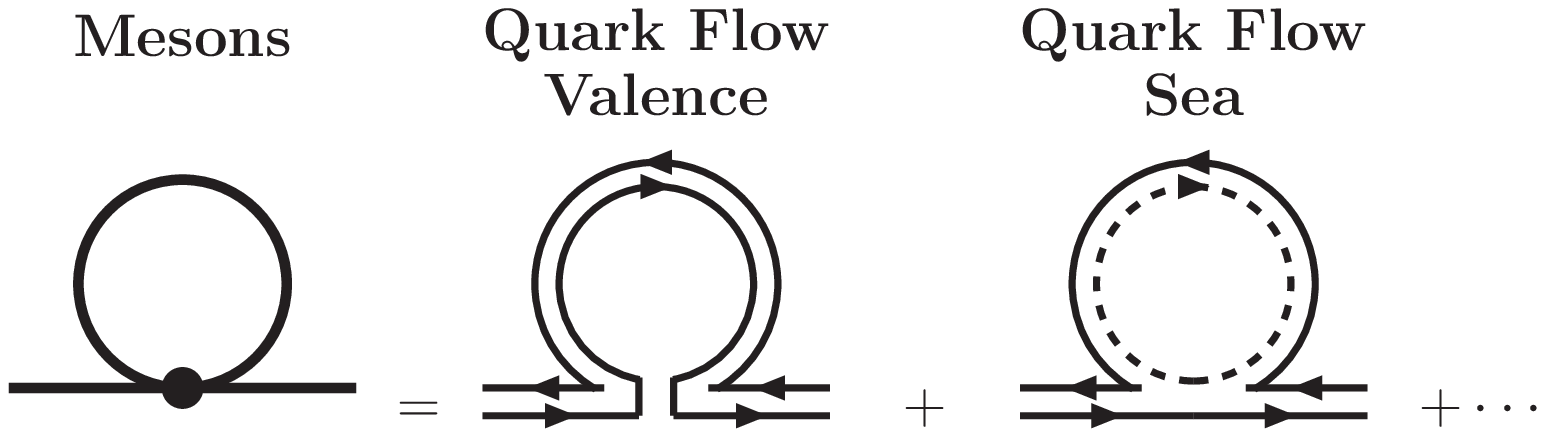}
\caption{\label{figPQ} An example of a meson (thick line)
loop diagram (left)
with the quark diagrams (right), there are gluons all over in the right
diagrams. The Valence (arrow line) and Sea (dash arrow line) quarks
can be treated independently in lattice QCD.}
\end{figure}

In the case of lattice QCD we need to extend this. One 
distinguishes between valence and sea quarks and they can be treated
independently. A simple example of how a meson loop contains quark loops
is shown in Fig.~\ref{figPQ}. In general it is much easier to
get to light valence quarks than to light sea quarks.
A more extended discussion can be found in \cite{BG,Sharpe}.
The version of ChPT with sea and valence quarks separately is called
Partially Quenched ChPT (PQChPT).

\section{Two Loop: General}

The Lagrangians needed at the first three orders, $p^2$, $p^4$
and $p^6$, are known and the number of parameters and their notation
are shown in Tab.~\ref{tabLag}.
\begin{table}
\begin{tabular}
{c@{$\quad\quad\quad$}cc@{$\quad\quad\quad$}cc@{$\quad\quad\quad$}cc}
\hline
      & \multicolumn{2}{c}{ 2 flavor} & \multicolumn{2}{c}{3 flavor} &
\multicolumn{2}{c}{ 3+3 PQChPT}\\
\hline
$p^2$ & $F,B$ & 2 & $F_0,B_0$ & 2 &  $F_0,B_0$ &  2 \\
$p^4$ & $l_i^r,h_i^r$ & 7+3 & $L_i^r,H_i^r$ & 10+2 & 
      $\hat L_i^r,\hat H_i^r$ &  11+2 \\
$p^6$ & $c_i^r$ & 53+4 & $C_i^r$ & 90+4 &  $K_i^r$ &
       112+3\\
\hline
\end{tabular}
\caption{\label{tabLag}
The number of free parameters and their names for the first
three orders in mesonic ChPT. The numbers refer to the number of
low energy constants and unmeasurable (high energy) constants.}
\end{table}
The parameters at order $p^2$ go back very long time.
The classification and infinities at
 $p^4$ can be found in \cite{GL1,GL2} and at $p^6$
in \cite{BCE1,BCE2}. The knowledge of the infinities provides a very useful
check on explicit calculations.

The replica method allows to obtain PQChPT from the $n_F$ flavor case
of \cite{BCE1,BCE2}. Important is that the 3 flavor case is a limit
of the 3+3 PQChPT one. Determining the parameters in PQChPT allows to
obtain those of 3 flavor ChPT via the Cayley-Hamilton relations of
\cite{BCE1}.
 
\section{Two Loop: 2 Flavours}

In this case most calculations have already been performed.\\
\parbox[t]{0.47\textwidth}{\topsep=0cm\partopsep=0cm
$\bullet$ $\gamma\gamma\to\pi^0\pi^0$ \cite{BGS}.\\
$\bullet$
$\gamma\gamma\to\pi^+\pi^-$, $F_\pi$, $m_\pi$ \cite{Burgi}.\\
$\bullet$
$\pi\pi$-scattering, $F_\pi$, $m_\pi$ \cite{BCEGS}.
}
\parbox[t]{0.47\textwidth}{\topsep=0cm\partopsep=0cm
$\bullet$
$F_{V\pi}(t)$, $F_{S\pi}$ \cite{BCT}.\\
$\bullet$
$\pi\to\ell\nu\gamma$ \cite{BT1}.
}
Here there is  in general a rather good convergence and it turns out that
for many threshold quantities the values of the $c^i_r$ are not
numerically important. These calculations are now often combined with
dispersive methods for very precise theoretical predictions.
An example is the full description of $\pi\pi$ scattering \cite{CGL}.

\section{Two Loop: PQChPT}

This subject is only beginning. The charged pion mass for
the case of all valence masses equal and all 3 sea quark masses equal
is published \cite{BDL} and the result for the decay constant in this case
will be published soon. Planned future work includes all the necessary mass
combinations as well as the two sea quark case.

\begin{figure}
\includegraphics[width=0.49\textwidth]{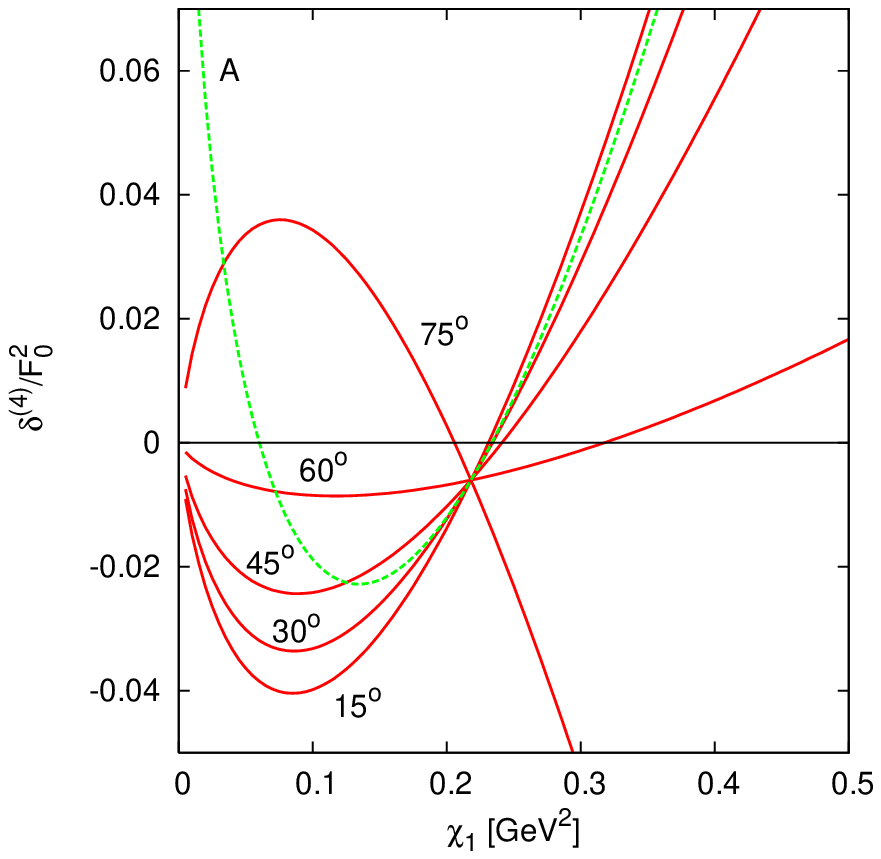}
\includegraphics[width=0.49\textwidth]{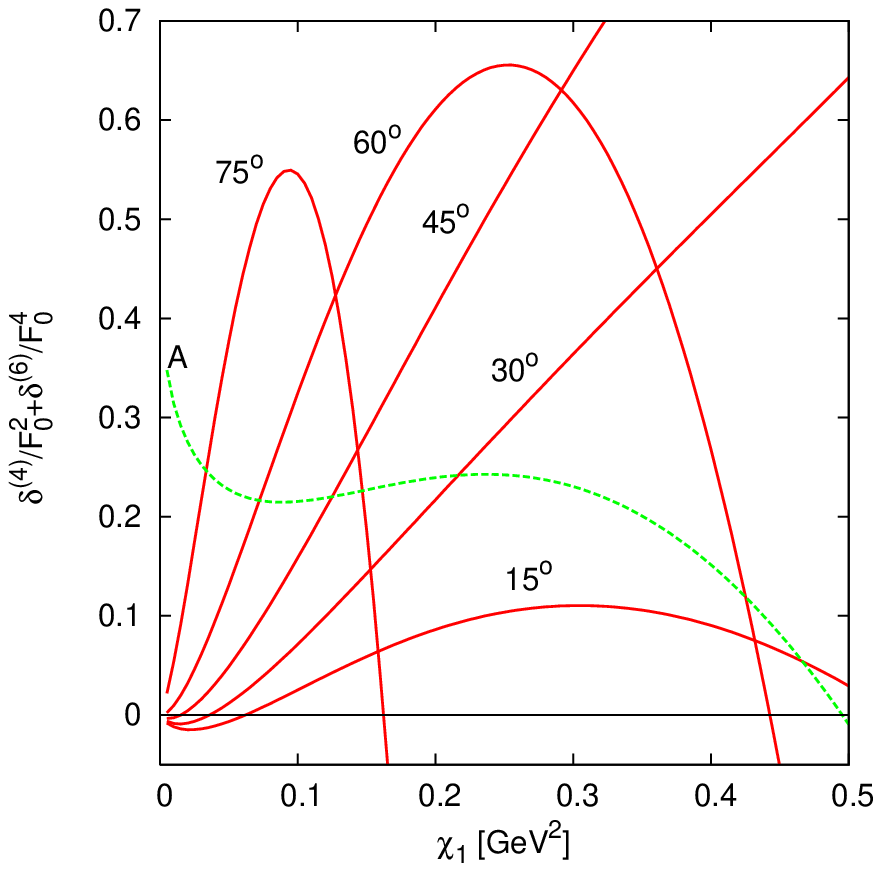}
\caption{\label{figPQresults} First results from PQChPT at
two-loop level \cite{BDL}. Shown is the relative correction to
the lowest order mass squared at order $p^4$ (left) and $p^4+p^6$ (right).
The valence quark mass is given as $\chi_1=2 B_0 m_{qV}$
and the sea quark mass as $\chi_4=2 B_0 m_{qS}$ with 
$\chi_4 = \chi_1\, \tan \theta$ (curves labeled with angle $\theta$)
or $\chi_4=0.125$~GeV$^2$ (A). $\chi_i$ corresponds to the lowest order
meson mass squared. The boundary for ChPT is reached for some mass squared
 around 0.3~GeV$^2$}
\end{figure}

In the actual calculations heavy use is made of the symbolic manipulation
program FORM \cite{FORM}. The major problem is simply the sheer size of
the expressions involved in the more general mass case due to the appearance
in PQChPT of many partial fractions of differences of quark masses.
An example of results is shown in Fig.~\ref{figPQresults}.

\section{Two Loop: 3 Flavor Overview and $\pi\pi$, $\pi K$ scattering}

Many calculations have been performed here. A (to my knowledge) complete
list is given below,
where the quantities which are determined from this calculation
are shown in brackets. The symbol $\Pi_{IJM}$ means a two-point Green
function of currents $I$ and $J$ with the quantum numbers of meson $M$.\\
\parbox[t]{0.60\textwidth}{
$\bullet$ 
$\Pi_{VV\pi}$, $\Pi_{VV\eta}$ \cite{GK1,ABT1}\\
$\bullet$
$\Pi_{VVK}$ \cite{ABT1,DK}\\
$\bullet$
$\Pi_{AA\pi}$, $\Pi_{AA\eta}$, $F_{\pi,\eta}$, $m_{\pi,\eta}$,
\cite{ABT1,GK2}\\
$\bullet$
$\Pi_{SS}$ \cite{Moussallam} ($L_4^r,L_6^r$)\\
$\bullet$
$\Pi_{VVK}$, $\Pi_{AAK}$, $F_K$, $m_K$ \cite{ABT1}\\
$\bullet$
$K_{\ell4}$ \cite{ABT2} ($L_1^r,L_2^r,L_3^r$)\\
$\bullet$ $F_{\pi,K,\eta}$, $m_{\pi,K,\eta}$  $(m_u\ne m_d)$
\cite{ABT3} ($L_{5,7,8}^r,m_u/m_d$)
}
\parbox[t]{0.39\textwidth}{
$\bullet$
$F_{V\pi}$, $F_{VK^+}$, $F_{VK^0}$  \cite{PS,BT2}
($L_9^r$)\\
$\bullet$
$K_{\ell3}$ \cite{PS,BT3} ($V_{us}$)\\
$\bullet$
$F_{S\pi}$, $F_{SK}$ \cite{BD} ($L_4^r,L_6^r$)\\
$\bullet$
$K,\pi\to\ell\nu\gamma$ \cite{GHW} ($L_{10}^r$)\\
$\bullet$
$\pi\pi$ \cite{BDT1}\\
$\bullet$
$\pi K$ \cite{BDT2}
}\\
We now perform a general fit to experiment and check how well
the whole system works and determine as many parameters as possible.
First we need to identify a series of basic inputs to determine
most of the parameters. The procedure is described in detail in
\cite{ABT2,ABT3}. The actual inputs used are
\begin{itemize}
\item 
$K_{\ell4}$: $F(0)$, $G(0)$, $\lambda$ from the E865 BNL experiment
\cite{Pislak}.
\item $m^2_{\pi^0}$, $m^2_\eta$, $m_{K^+}^2$, $m_{K^0}^2$ 
Electromagnetic corrections including quark mass effects\cite{Dashen}.
\item $F_{\pi^+} = 92.4$~MeV, $F_{K^+}/F_{\pi^+} = 1.22\pm0.01$,
 $2 m_s/(m_u+m_d) = 24$ or $26$.
\item $L_4^r, L_6^r$.
\item The $C_i^r$ as estimated using single resonance approximation.
\end{itemize}
These fits are performed varying the resonance input used for the $C_i^r$
by varying the size by an overall factor of two and the
scale at which the saturation is applied. 
The final fitting result with errors
is shown for some typical inputs in Tab.~\ref{tabfitLi}. The $C_i^r$
can be determined from experiment in several cases. When the dependence
is purely kinematical, e.g. curvature of a form-factor, it works
reasonably well \cite{BCT,BT2}. For those with a mixed quark mass-kinematical
dependence, e.g. the slope of $f_+(t)$ in $K_{\ell 3}$ it works OK \cite{BT3}.
The remaining ones are difficult to estimate, the question here is
what type of scalar to use, we know it is not the sigma \cite{BCT,BD}
but what else is an open question. 

\begin{figure}
\includegraphics[width=0.9\textwidth]{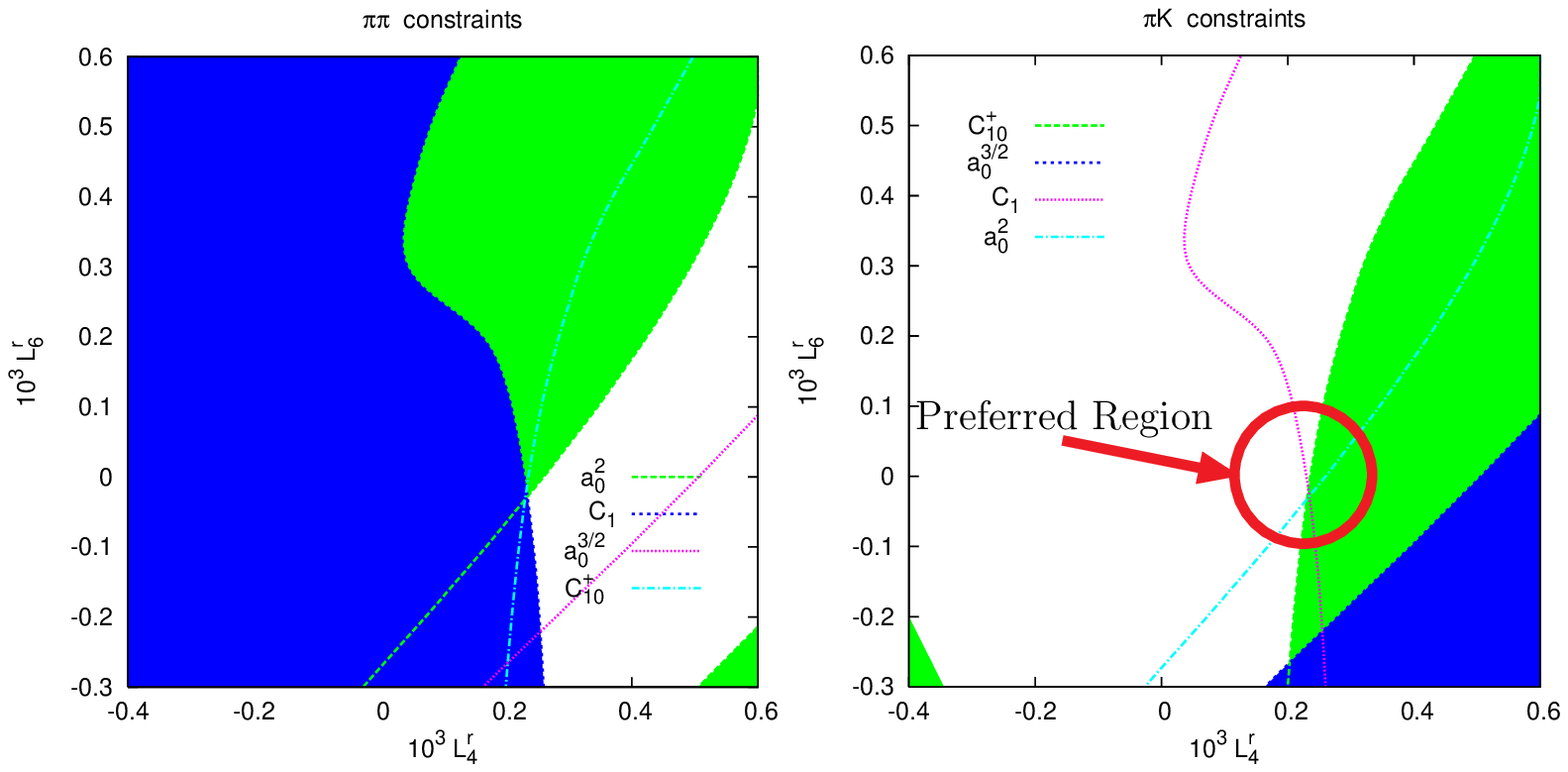}
\caption{\label{figconstraints} The constraints on $L_4^r,L_6^r$
from $\pi\pi$ (left) and $\pi K$ (right) scattering \cite{BDT1,BDT2}.}
\end{figure}

The whole procedure is repeated for a range of input values
of $L_4^4,L_6^r$. Some examples of the fits are given in Tab.~\ref{tabfitLi}
where we quote the main fit (labeled fit 10), the same one but at $p^4$
order rather than $p^6$ as well as two with $L_4^4,L_6^r\ne0$. Fit B
is one where all scalar form factors behave nicely \cite{BD} while fit D
is the one where the threshold parameters for $\pi\pi$ and $\pi K$ scattering
are well fitted as discussed in more detail in \cite{BDT2}.
Fig.~\ref{figconstraints} shows the constraints
from $\pi\pi$ and $\pi K$ scattering as well as the region of fit D.
\begin{table}
\begin{tabular}{ccccc}
                & fit 10 & same $p^4$ & fit B & fit D\\
\hline
$10^3 L_1^r$ & $0.43\pm0.12$ & $0.38$ & $0.44$ & $0.44$\\
$10^3 L_2^r$ & $0.73\pm0.12$ & $1.59$ & $0.60$ & $0.69$\\
$10^3 L_3^r$ & $-2.53\pm0.37$ & $-2.91$ &$-2.31$&$-2.33$\\
$10^3 L_4^r$ & $\equiv0$    & $\equiv 0$& $\equiv0.5$ & $\equiv0.2$\\
$10^3 L_5^r$ & $0.97\pm0.11$& $1.46$ & $0.82$ & $0.88$\\
$10^3 L_6^r$ & $\equiv0$    & $\equiv 0$& $\equiv0.1$ & $\equiv0$\\
$10^3 L_7^r$ & $-0.31\pm0.14$&$-0.49$ & $-0.26$ & $-0.28$\\
$10^3 L_8^r$ & $0.60\pm0.18$ & $1.00$ & $0.50$ & $0.54$\\
\hline
$2 B_0 \hat m/m_\pi^2$ & 0.736 & 0.991 & 1.129 & 0.958\\
$m_\pi^2$: $p^4,p^6$    & 0.006,0.258 & 0.009,$\equiv0$ & $-$0.138,0.009 &
          $-$0.091,0.133\\
$m_K^2$: $p^4,p^6$    & 0.007,0.306 & 0.075,$\equiv0$ & $-$0.149,0.094 &
          $-$0.096,0.201\\
$m_\eta^2$: $p^4,p^6$    & $-$0.052,0.318 & 0.013,$\equiv0$ & $-$0.197,0.073 &
          $-$0.151,0.197\\
$m_u/m_d$    & 0.45$\pm$0.05 & 0.52 & 0.52 & 0.50\\
\hline
$F_0$ [MeV]          & 87.7 & 81.1 & 70.4 & 80.4 \\
$\frac{F_K}{F_\pi}$: $p^4,p^6$ & 0.169,0.051 & 0.22,$\equiv0$ & 0.153,0.067 &
   0.159,0.061\\
\end{tabular}
\caption{\label{tabfitLi} The results of the fits for several input values
of $L_4^r,L_6^r$, see text.}
\end{table}

\section{Two Loop: $K_{\ell 3}$ and \mbox{$V_{\lowercase{us}}$}}

The CKM matrix element $V_{us}$ is a fundamental parameter we like to determine
as precisely as possible. In addition, the relation  $|V_{ud}|^2+|V_{us}|^2=1$
is broken at about the two sigma level by the experimental values quoted
in the PDG 2002\cite{PDG2002}. It is thus important to update both
theory and experiment. The theory \cite{LR} has been updated in
several ways. The photonic corrections have been properly calculated
in the modern ChPT language \cite{Cirigliano} and the form factors
$f_+(t)$, $f_-(t)$ are known to $p^6$ in ChPT \cite{PS,BT3}.

The experiments have been mainly analyzed by assuming a linear parameterization
of the form-factor. This is not sufficient as was pointed out in \cite{BT3}.
The value of $|f_+(0)V_{us}|$ changes by 0.9\% (0.6\%),
data from \cite{CPLEAR} (\cite{PSE246}), when a linear fit is used
as compared with a quadratic one where the curvature is determined
from ChPT \cite{BT3}. The newer experiments \cite{KTeV,ISTRA}
have now detected
the curvature and find values in agreement with the ChPT prediction of
\cite{BT3}. The measured branching ratio in both the neutral and
charged channel has also increased, \cite{KTeV,BNL,KLOE}.
Both combined lead to an increase in
the value of $|f_+(0) V_{us}|$ and using the value of $f_+(0)$ from \cite{LR}
the unitarity problem is resolved.

The remaining problem is to accurately predict the value of $f_+(0)$.
In \cite{LR} one-loop ChPT and a quark model estimate of the higher orders
were used. The $p^6$ calculation gives loop corrections
of about one \% to the one loop result \cite{PS,BT3}. The full
analysis including updated values of all the inputs is in \cite{BT3}
but a more important point was made in \cite{BT3} as well.
The $p^6$ low energy constants that appear in the value of $f_+(0)$
can be determined experimentally from the scalar form factor
$f_0(t)$ since \cite{BT3}
\begin{eqnarray}
f_0(t) &=& 1-\frac{8}{F_\pi^4}{\left(C_{12}^r+C_{34}^r\right)}
\left(m_K^2-m_\pi^2\right)^2
+8\frac{t}{F_\pi^4}{ \left(2C_{12}^r+C_{34}^r\right)}
\left(m_K^2+m_\pi^2\right)
\nonumber\\&&
+\frac{t}{m_K^2-m_\pi^2}\left(F_K/F_\pi-1\right)
-\frac{8}{F_\pi^4} t^2 { C_{12}^r}
+\overline\Delta(t)+\Delta(0)\,.
\end{eqnarray}
In this equation everything is 
known except the values of $C_{12}^r$
and $C_{34}^r$, 
and correlations between $F_K/F_\pi$ and $V_{us}$\cite{knecht}. 
Since $f_+(0)=f_0(0)$ a measurement of the slope
and curvature of the scalar form factor allows to determine
$f_+(0)$. Dispersion theory allows to relate
some of these quantities as well. A first analysis \cite{JOP}
leads to an estimate that essentially cancels the pure loop contribution
yielding a very small total $p^6$ correction.


\section{Conclusions}

The two flavor case in ChPT at two loops is an almost finished subject.
The three flavor case is in progress. Many calculations have been done
and things seem to work but convergence is sometimes slow.
$L_4^r,L_6^r$ are nonzero but reasonable for large $N_c$ expectations.
An example of clean predictions even with the many $p^6$ constants is
the case of $K_{\ell3}$ decays and the determination of $V_{us}$.
The partially quenched case, relevant for lattice calculations, is
just at its beginnings at two loop order.

\begin{theacknowledgments}

Supported in part by the Swedish Research Council
and the EU TMR
network, Contract No. HPRN-CT-2002-00311  (EURIDICE).

\end{theacknowledgments}

\end{document}